\documentclass[apl, bibnotes, showpacs, preprintnumbers, amsmath, amssymb, superscriptaddress]{revtex4-1}

\usepackage{graphicx}
\usepackage{dcolumn}
\usepackage{bm}
\usepackage{color}
\usepackage{colortbl}
\usepackage{url}
\usepackage{setspace}
\usepackage {epsf}
\usepackage {epsfig}
\usepackage{times}
\usepackage {caption}
\usepackage {graphicx}
\usepackage {graphics}
\usepackage {array}

\begin{document}

\bibliographystyle{apsrev}

\title{Sidewall depletion in nano-patterned LAO/STO heterostructures}

\author{M. Z. Minhas}
\affiliation{Institut f\"{u}r Physik, Martin-Luther-Universit\"{a}t Halle-Wittenberg, Von-Danckelmann-Platz 3, D-06120 Halle, Germany}
\author{H. H. Blaschek}
\affiliation{Institut f\"{u}r Physik, Martin-Luther-Universit\"{a}t Halle-Wittenberg, Von-Danckelmann-Platz 3, D-06120 Halle, Germany}
\author{F. Heyroth}
\affiliation{Interdisziplin\"{a}res Zentrum f\"{u}r Materialwissenschaften,Martin-Luther-Universit\"{a}t Halle-Wittenberg,Heinrich-Damerow-Str. 4, 06120 Halle, Germany}
\author{G. Schmidt}
\email[Correspondence to G. Schmidt: ]{georg.schmidt@physik.uni-halle.de}
\affiliation{Institut f\"{u}r Physik, Martin-Luther-Universit\"{a}t Halle-Wittenberg, Von-Danckelmann-Platz 3, D-06120 Halle, Germany}
\affiliation{Interdisziplin\"{a}res Zentrum f\"{u}r Materialwissenschaften,Martin-Luther-Universit\"{a}t Halle-Wittenberg,Heinrich-Damerow-Str. 4, 06120 Halle, Germany}

\begin{abstract}
We report the fabrication of nanostructures from the quasi-two-dimensional electron gas (q2DEG) formed at the LaAlO$_{3}$/ SrTiO$_{3}$ (LAO/STO) interface. The process uses electron beam lithography in combination with reactive ion etching. This technique allows to pattern high-quality structures down to lateral dimensions as small as $100$nm while maintaining the conducting properties without inducing conductivity in the STO substrate. Temperature dependent transport properties of patterned Hall bars of various widths show only a small size dependence of conductivity at low temperature as well as at room temperature. The deviation can be explained by a narrow lateral depletion region. All steps of the patterning process are fully industry compatible.
\end{abstract}
\maketitle
The discovery of an electron gas at the interface between the two band insulators LaAlO$_{3}$ (LAO) and SrTiO$_{3}$ (STO)\cite{Ohtomo2004,Thiel2006} has initiated a huge effort to study this interface in detail. Besides the interface conductivity also other interesting properties have been reported such as induced ferromagnetism at the interface between the two non-magnetic insulating perovskites\cite{Brinkman2007} and superconductivity below $200$\,mK. \cite{Reyren2007,Li2011} To understand the origin of the electrical conductivity different models have been proposed; polar discontinuity at the interface,\cite{Nakagawa2006} formation of oxygen vacancies in the STO substrate during the growth of LAO on STO substrate,\cite{Siemons2007,Kalabukhov2007,Chen2011} or La-Sr intermixing at the interface. \cite{Willmott2007} Due to the above mentioned properties the LAO/STO interface has not only become a model system to study the fundamental physics of strongly correlated electronic systems but also a candidate for future multifunctional oxide electronics. \cite{Forg2012,Hosoda2013}

Still one challenge has to be faced. For both, using the material in nanopatterned quantum transport devices or employing it in state of the art field effect transistor based nanoelectronics a reliable and reproducible patterning technique is essential to achieve lateral insulation between devices. Common dry etching techniques like Ar ion milling are of only limited use for LAO/STO structures, because exposure to the ion beam creates a highly conducting layer on the SrTiO$_{3}$ substrate surface at room temperature as well as at cryogenic temperatures.\cite{Reagor2005,Kan2005,Gross2011} This highly conducting layer complicates any transport experiment in the patterned structures and makes device applications virtually impossible. Up to now, however, a number of indirect patterning processes have been reported. One successful approach for patterning makes use of the particular thickness dependence of the interface conductivity, which sets in when the thickness of the LAO layer exceeds $4$ unit cells (u.c.) only. In this approach of Schneider et al.\cite{Schneider2006} $2$u.c of epitaxial LAO were grown on STO substrate. Subsequently UV-lithography was done and amorphous LAO was deposited which was then patterned by lift off. Further deposition of LAO under suitable growth conditions led to a continuation of epitaxial growth limited to those places where the epitaxial LAO surface was still open while in places covered by amorphous material crystalline growth was inhibited. In places where epitaxial growth occurred the condition for a conducting interface was fulfilled. Thus a patterning of the conductivity without physical patterning of the interface was achieved. In another approach Banjerjee et al.\cite{Banerjee2012} used an $AlO_{x}$ based hard mask and epitaxial lift-off, leaving the desired areas of LAO/STO with insulating STO in between. In their approach, however, no nanopatterning was demonstrated and the STO substrate was exposed to lithography and solvents before the deposition of LAO. While these two processes used epitaxial growth on the pre-patterned surfaces there were also attempts to locally modify the conductivity in large area LAO/STO layers.

Quite frequently conducting atomic force microscope (AFM) probes were used to create and erase conducting nanoscale structures at the LAO/STO interface. Quantum size effect could be demonstrated, \cite{Cen2008,Cen2009} however, the disadvantage of this technique is that the structures are not stable over time at ambient conditions and it is difficult to pattern large area devices or even integrated circuits by using this technique. Recently, low energy Ar-ion beam irradiation was used to pattern the quasi-two-dimensional electron gas (q2DEG) at the LAO/STO interface with the combination of optical and electron beam lithography and subsequent ion beam irradiation. \cite{PaoloAurino2013} The process relies on the fact that the decrease in interface conductivity under ion radiation due to lattice damage is faster than the increase of the substrate conductivity, leaving only a small process window for pattern fabrication. Lateral dimensions of $50$nm were achieved, however, at the cost of an unexplained increase in resistivity of almost one order of magnitude. No real physical nano-patterning of large area LAO/STO has been reported so far because of the issue of ion induced conductivity of the STO surface.
\begin{figure}
\begin{center}
\includegraphics[width=17cm]{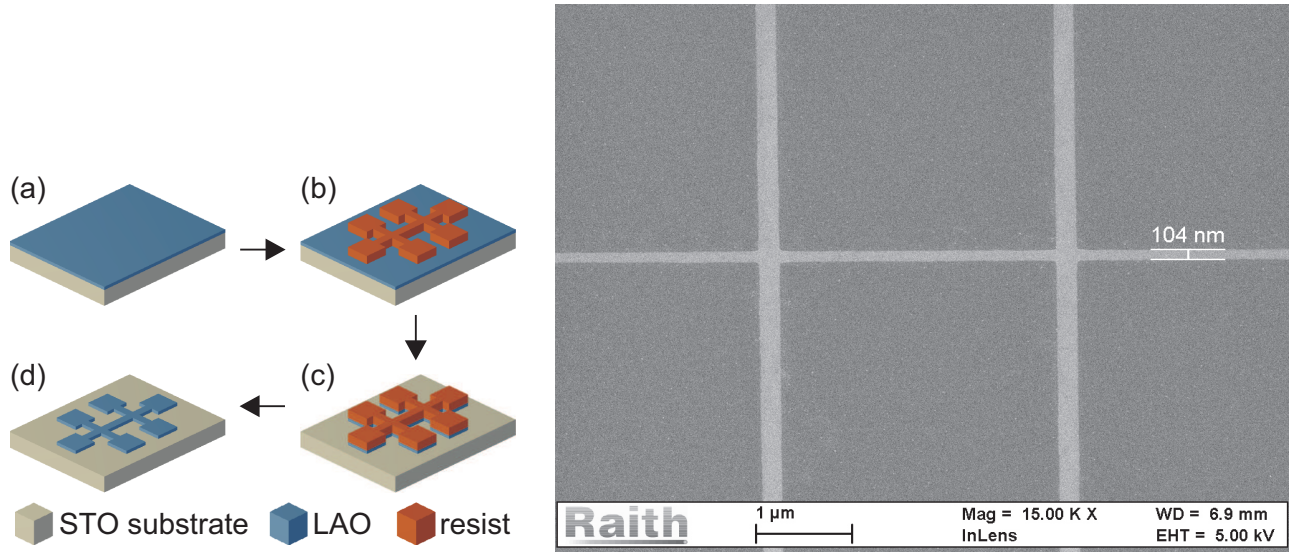}
\caption{Left: Schematic diagram showing the process used to pattern the q2DEG down to nanometer scale. (a) The LAO film is deposited on the $TiO_2$-terminated STO substrate. (b) Resist is deposited and patterned by e-beam lithography. (c) Dry etching is performed down to the STO substrate. (d)After removal of the resist, the patterned LAO structure remains on the STO. Right: Scanning electron micrograph of part of a $100\,nm$ wide Hall bar with approx. $500\,nm$ wide contact leads at the top and bottom.\label{Fig1}}
\end{center}
\end{figure}

Here we present a reliable technique to physically pattern the q2DEG down to lateral dimensions as small as $100$nm while maintaining its conducting properties, however, without the problem of an ion induced substrate conductivity.

We use LAO layers (6.u.c) grown by pulsed laser deposition (PLD) from a single crystal LAO target on $TiO_{2}$-terminated STO (001) substrates.\cite{Kawasaki1994,Koster1998,Huijben2006} Oxygen is used as a background gas at a pressure of $10^{-3}$ mbar during the deposition. The substrate temperature during deposition is 850$^{\circ}$C. Laser fluence and pulse frequency are kept at 2 J/cm$^{2}$ and 2Hz, respectively, during the deposition. To monitor the layer thickness up to unit cell level in situ reflection high-energy electron diffraction (RHEED) is used during the growth.\cite{Rijnders1997} After deposition the samples are slowly cooled down to room temperature while the oxygen pressure is maintained.

For the patterning a thin film of novolack based image reversal resist is deposited by spin coating (see Fig. \ref{Fig1}, left). Subsequently the sample is exposed by electron beam (e-beam) lithography. The resist we use is also suitable for high resolution optical lithography allowing for direct transfer of the process to industrial lithography tools. The exposed pattern consists of Hall bars with different respective nominal width between $100$ and $500 nm$ including large area bond pads. After development reactive ion etching is performed in an Inductively coupled plasma reactive ion etching system (ICP-RIE, Oxford Plasmalab $100$). The etching process uses a pressure of $5$\,mTorr of $BCl_{3}$. The plasma is excited with a combination of RIE and ICP at a total power of $1430$W. The sample temperature is kept at 5$^{\circ}$C by helium backside cooling. Using these parameters we achieve an etch rate of $13\pm3\,nm/min$. We chose a process time of $19\,s$ in order to completely remove the LAO layer. After the etching the resist is removed using N-Ethylpyrrolidon at 60$^{\circ}$C for $3$ hours. The resulting patterned structures are stable at ambient conditions. The Hall bars are bonded using Al wire for electrical transport measurements. The bonds are placed directly on the LAO without additional metallization.
Fig. \ref{Fig1} (right) shows a $100\,nm$ wide bar with approx. $500\,nm$ wide contact leads.

\begin{figure}
\begin{center}
\includegraphics[width=8.5cm]{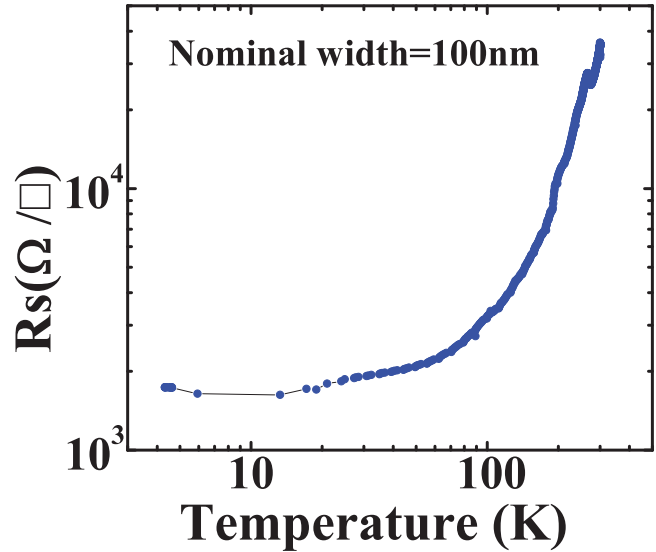}
\caption{Temperature dependent sheet resistance of the q2DEG in Hall bar geometry with a nominal width of 100nm.\label{Fig2}}
\end{center}
\end{figure}
\begin{figure}
\begin{center}
\includegraphics[width=17cm]{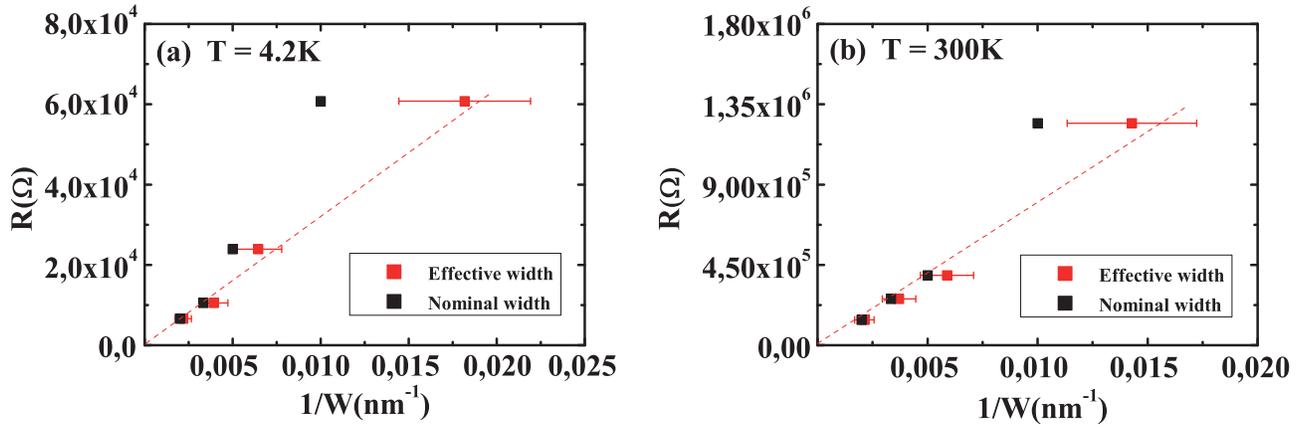}
\caption{Resistance of the q2DEG in Hall bar geometry plotted as a function of the inverse effective (red) and nominal (black) width of the Hall bars.  (a) Measured at 4K, (b) measured at 300K.\label{Fig3}}
\end{center}
\end{figure}
\begin{figure}
\begin{center}
\includegraphics[width=17cm]{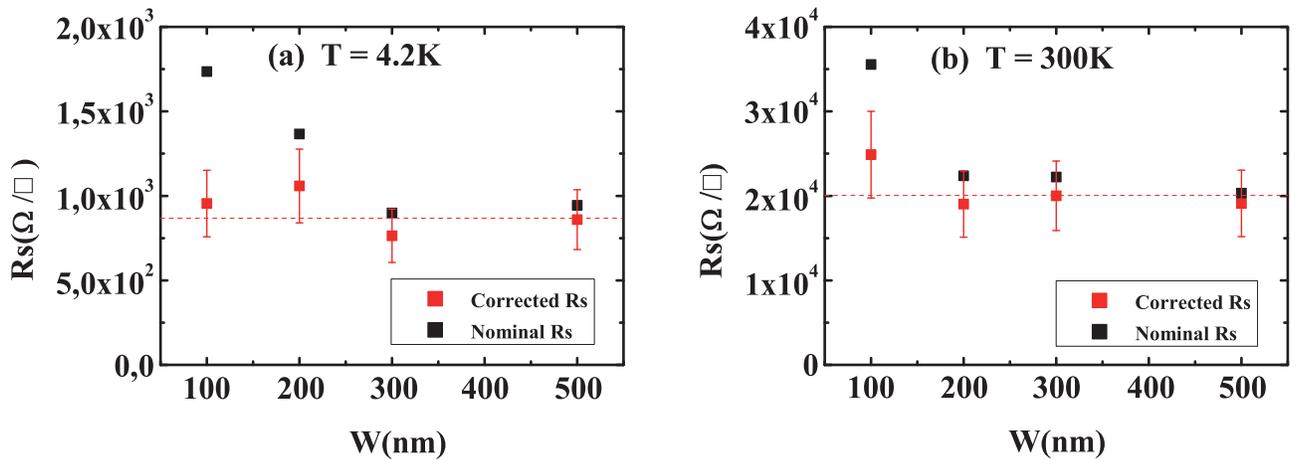}
\caption{Sheet resistance of the q2DEG in Hall bar geometry plotted as a function of width. When the effective width is used, the sheet resistance is constant within th error bars. (a) Measured at 4K, (b) measured at 300K.\label{Fig4}}
\end{center}
\end{figure}

Electrical transport measurements are carried out in a $^{4}$He bath cryostat in the temperature range of $4.2-300$K. We are using DC measurements with a voltage drop over the sample of approximately $5-7$mV. The voltage drop over the sample and the reference resistor, respectively, as well as the Hall voltage are measured using high impedance difference amplifiers and an Agilent $34420$A nanovoltmeter. All fabricated Hall bars show metallic behavior down to $4.2$K (see Fig. \ref{Fig2}). The area between the structures is insulating at room as well as at low temperature. Fig. \ref{Fig3}(a and b) shows the resistance as a function of the Hall bars inverse nominal width at room as well as at low temperature. Although the plots show almost straight lines as expected for constant conductivity, there is a notable increase in sheet resistance with decreasing dimension of the structures.
This effect can be seen more precisely when the sheet resistance is plotted as a function of width (see Fig. \ref{Fig4}). For a perfect etching process the sheet resistance should be constant and independent from the width of the Hall bars. In our results, however, comparing the temperature dependent resistivity for large area sample and micro Hall bars of different widths yields the following: For large area samples the sheet resistance is approximately $12$K$\Omega/\square$ at room temperature and $0.2$K$\Omega/\square$ at $4.2$K. In the Hall bars we observe a slight increase in sheet resistance with decreasing width.  This increase, however, is much weaker than the one observed by Aurino et al.\cite{PaoloAurino2013}. In order to understand the origin of the effect we analyze the dependence of the resistance on the width of the structures in more detail. While an overall increase of the resistivity of the nanopatterned samples does not fit the experimental results, side wall damage can explain the observations.

During the etching process, the sidewalls are exposed to the etchant and the material is damaged to a certain extent, mainly depending on the etching time. As the etching time is constant for all structures, the depth of the damage in the crystal is also identical for all Hall bars. This damage can result in a non-conducting depletion region with constant width for all structures. If this assumption is true introducing an 'effective' width for the Hall bars in which a constant value 'x' is subtracted from the nominal width should yield a constant sheet resistance for all Hall bars. For the sake of simplicity we assume an infinite resistance for the depletion region. The sidewall depletion length x can be found by fitting Hall bars of different width to the formula:
\begin{equation}
R_{total} = \frac{\rho_{1}\rho_{2}L}{\rho_{2}t_{1}w-x(\rho_{2}t_{1}-\rho_{1}t_{2})}
\end{equation}

where $\rho_{1}$ and $\rho_{2}$ are electrical resistivity and 'w' and 'x' is width of Hall bar and depletion region respectively. It should be noted that the width of the depletion region is expected to depend on the temperature and thus for the fitting procedure two different respective depletion regions must be assumed for T=$4.2\,K$ and for the room temperature. This assumption takes into account that defects introduced by the etching may be traps whose occupation depends on the energy level and thus on the temperature. Based on this analysis we get a best fit for a side wall depletion of approximately $15\,nm$ on each side of structure at room temperature and $20\,nm$ at $4.2\,K$. Fig. \ref{Fig4} shows the sheet resistance plotted over the nominal width once calculated using the nominal width and once calculated using the effective width. Obviously the value which is obtained using the effective width is constant within the error bars of approximately $20$\%\ at $4.2\,K$ as well as at room temperature confirming the validity of our model.

In conclusion we have demonstrated that it is possible to pattern the electron gas at the LAO/STO interface using e-beam lithography in combination with dry etching. The process preserves the conductivity of the electron gas but leaves the substrate insulating and thus can be used for the lateral insulation of electrical devices. Our analysis shows that a small depletion region is induced at the lateral surfaces which reduces the effective width of the structures. Nevertheless, the resolution is good enough to access device dimensions well below $100\,nm$. Using a Novolac based resist which is also suitable for optical lithography and ICP and RIE based reactive chemical dry etching the process is industry compatible and it is expected that more detailed optimization of the process conditions can even further reduce the already small surface damage.

\begin{acknowledgements}
This work was supported by the European Commission in the project IFOX under grant agreement NMP3-LA-2010-246102 and by the DFG in the SFB 762.
\end{acknowledgements}


\end{document}